\documentclass[10pt]{article}
\usepackage{tmlr}

\usepackage{cite}
\usepackage{hyperref}
\usepackage{url}
\usepackage{amsmath,amssymb,graphicx,hyperref,booktabs}
\usepackage{algorithm}
\usepackage{algpseudocode}
\usepackage{tikz}
\usepackage{multirow}
\usepackage{booktabs}
\usepackage{array}
\usepackage{placeins}
\usepackage{natbib}
\usetikzlibrary{shapes.geometric, arrows.meta, positioning, fit, backgrounds, calc}

\title{Systemic Risk Radar: A Multi-Layer Graph Framework for Early Market Crash Prediction}

\begin{document}

\maketitle

\begin{abstract}
Financial crises emerge when structural vulnerabilities accumulate across sectors, markets, and investor behavior. Predicting these systemic transitions is challenging because they arise from evolving interactions between market participants, not isolated price movements. We present Systemic Risk Radar (SRR), a framework that models financial markets as multi-layer graphs to detect early signs of systemic fragility and crash-regime transitions.

We evaluate SRR across three major crises: the Dot-com crash, the Global Financial Crisis, and the COVID-19 shock. Our experiments compare snapshot GNNs, a simplified temporal GNN prototype, and standard baselines (logistic regression and Random Forest). Results show that structural network information provides useful early-warning signals compared to feature-based models alone.

This correlation-based instantiation of SRR demonstrates that graph-derived features capture meaningful changes in market structure during stress events. The findings motivate extending SRR with additional graph layers (sector/factor exposure, sentiment) and advanced temporal architectures (LSTM/GRU or Transformer encoders) to better handle diverse crisis types.

\end{abstract}

\section{Introduction}
Financial markets are interconnected. Price formation, liquidity, and volatility reflect sector-wide and cross-asset feedback effects. Large market declines over the last two decades—the Dot-com crash (2000--2002), the global financial crisis (2008--2010), and the COVID-19 crash (2020)—were not triggered by single shocks. They emerged from accumulated fragility: rising correlations, leverage amplification, deteriorating liquidity, and cross-market contagion.

Most market prediction tools focus on individual assets or sector-wise indicators. Factor models, volatility indexes, and macroeconomic indicators struggle to capture systemic interactions. They predict whether specific assets or sectors will decline, not whether the broader market is approaching a systemic risk regime. This requires understanding the connectivity and temporal evolution of market structure.

We introduce a modeling approach based on a simple observation: systemic instability shows up in the structure of the market before it shows up in prices. When correlations spike, volatility clusters spread, and sectors synchronize, the system becomes vulnerable. These structural signals precede actual drawdowns and can warn of impending regime shifts.

Recent research has explored complementary approaches at finer levels of granularity. 
Structural price decomposition methods have been proposed to extract explainable market 
behavior at the individual instrument level, while multimodal frameworks have been used 
to identify manipulation and anomaly patterns by aligning market data with social signals 
\citep{neela2025spa, neela2025aimm}. 
However, these approaches operate primarily at the micro or event level and do not model 
how localized stress accumulates and propagates across the broader market system. 
SRR is designed to address this gap by modeling systemic risk as an emergent property of 
market-wide structure rather than isolated signals.

Past work shows that systemic fragility emerges through the build-up of network connectivity and amplification effects, not isolated price shocks ~\citep{battiston2012debtrank, mantegna1999hierarchical, haldane2011systemic}.
Traditional forecasting models miss these structural dependencies and network-mediated channels of contagion ~\citep{laloux2000random, acemoglu2015systemic}.

Recent work on contagion in financial networks further highlights how localized shocks can propagate system-wide through balance sheet and funding linkages ~\citep{glasserman2015contagion, elliott2014financial, gai2010contagion}.

\subsection{Motivation}
SRR addresses two problems. First, systemic risk is a macro-level phenomenon that single-variable predictors cannot explain. Second, traditional models do not exploit recent advances in graph neural networks, which can learn representations of evolving networks. Our key insight: model financial markets as dynamically evolving graphs, where nodes represent instruments and edges represent relationships like correlation, co-volatility, or sentiment co-movement.

\subsection{Gaps in Prior Approaches}
Prior work relies on historical time series, macro signals, or factor models. These methods do not capture multi-sector connectivity or allow representations to evolve over time. Even correlation matrices and rolling covariance compress market structure into static snapshots, losing temporal dynamics.

\subsection{Contributions}
This paper makes the following contributions:
\begin{itemize}
    \item We introduce a multi-layer graph representation for market states, capturing correlation, sector similarity, and sentiment connectivity.
    \item We propose a temporal GNN framework for modeling systemic risk evolution and implement a simplified prototype using GCN encoders with GRU-based sequence aggregation.
    \item We provide methods for interpretability and scenario simulation designed to help practitioners understand model decisions.
    \item We demonstrate that SRR provides meaningful early-warning behavior relative to baselines.
    \item We analyze how different stress regimes expose distinct structural limitations of correlation-based GNNs.
\end{itemize}

Unlike prior micro-level approaches, SRR explicitly targets system-wide risk dynamics and cross-asset propagation.

\subsection{Empirical Findings}
Preliminary empirical analysis shows that graph-based representations provide meaningful early-warning signals of systemic fragility. In particular, simple snapshot GNN models, a simplified temporal GNN prototype, and traditional baselines (logistic regression and Random Forest) capture shifts in correlation structure and volatility clustering during known crisis events. The evaluated temporal model uses only the correlation layer and short graph sequences, but still demonstrates that market topology carries predictive information. These results provide preliminary support for the SRR framework and motivate training the full multi-layer temporal architecture with richer inputs.

\section{Related Work}

Network-based contagion modeling has been widely studied in systemic risk research
~\citep{battiston2012debtrank, haldane2011systemic, acemoglu2015systemic}, while correlation
networks and random matrix theory have been used to characterize market structure and
instability ~\citep{mantegna1999hierarchical, laloux2000random}. More recently, graph-based
and temporal neural architectures have been applied to financial prediction tasks
~\citep{huang2020financialgnn, feng2019temporalgnn}, but their use for macro-level systemic regime detection remains limited.

Systemic risk modeling spans early-warning systems, network-based contagion models, and deep learning for financial forecasting. Traditional early-warning systems use macroeconomic indicators like credit spreads, volatility indexes, and interest rate slopes. These identify broad stress but miss structural market dynamics.

Network models analyze contagion in interbank lending and credit networks through exposure networks, overlapping portfolios, and counterparty risk. Correlation networks proxy market connectivity, and rising correlation precedes instability. But correlation-only approaches produce a single market representation and miss other forms of connectivity.

Deep learning methods (RNNs, hybrid sentiment-market models) forecast volatility and stress but focus on asset-level dynamics, not systemic transitions.

Graph neural networks appear in portfolio optimization and risk propagation but temporal GNNs for systemic regime detection are rare. We know of no prior work combining heterogeneous multi-layer graphs with temporal sequence modeling for systemic risk prediction.

In parallel, the graph neural network literature has developed rapidly, from early graph convolutional networks ~\citep{kipf2017semi} and graph attention architectures ~\citep{velivckovic2018graph} to broader surveys of GNN methods ~\citep{wu2020comprehensive}. These models form a natural foundation for learning over financial network structures.

\section{Data and Feature Engineering}
We utilize publicly available market and macroeconomic data, including:

\begin{itemize}
    \item Equity index and sector ETF prices,
    \item Volatility indexes such as VIX (available but not used in current experiments),
    \item Treasury yields and credit spread proxies (future work),
    \item Sentiment and news-derived indicators (future work).
\end{itemize}

From these sources, we construct seven daily node-level features: daily returns, 20-day and 60-day rolling volatility, 20-day and 60-day drawdowns, and 10-day and 30-day momentum indicators. Macro overlays reflecting economic conditions are included as graph-level context features. Feature engineering is performed using rolling windows to ensure no future information is inadvertently incorporated.

Table~\ref{tab:datasets} summarizes the dataset. The crisis periods selected for evaluation---the Dot-com crash, global financial crisis, and COVID shock---represent structurally different regimes, testing SRR's generality.

\begin{table}[t]
\centering
\caption{Summary of crisis periods and dataset characteristics used in our experiments.}
\label{tab:datasets}
\begin{tabular}{lccc}
\toprule
\textbf{Crisis} & \textbf{Period} & \textbf{Trading Days} & \textbf{\# Stocks} \\\\
\midrule
Dot-com Bubble & 1998--2003 & 1{,}565 & 44 \\\\
Global Financial Crisis & 2006--2011 & 1{,}565 & 44 \\\\
COVID--19 Pandemic & 2018--2021 & 1{,}045 & 44 \\\\
\bottomrule
\end{tabular}
\end{table}

\section{Multi-Layer Graph Model}

Empirical work shows that changes in network topology and connectivity precede large market dislocations \citep{haldane2011systemic, fouque2013systemic}.

SRR's premise: systemic fragility is encoded in market structure, not isolated asset movements. We represent the market as a dynamically evolving multi-layer graph. Each layer corresponds to a different relationship type between financial assets, and each layer evolves over time as market conditions change.

\subsection{Graph Construction}
Let $G_t = (V, E_t)$ be the market graph at time $t$, where $V$ is the set of nodes representing instruments and $E_t$ is the set of edges capturing connectivity. We consider three primary edge types:

\begin{itemize}
    \item \textbf{Correlation Layer:} edges represent historical return correlations, computed over a sliding window.
    \item \textbf{Sector/Factor Layer:} edges link entities belonging to the same sector or factor exposure.
    \item \textbf{Sentiment Co-movement Layer (optional):} edges represent co-movement in social-media or sentiment signals.
\end{itemize}

Each layer is represented by a separate adjacency or edge set, allowing the model to disentangle different forms of connectivity.

Formally, the multi-layer graph is:
\[
\mathcal{G}_t = \{\mathcal{G}_t^1, \mathcal{G}_t^2, \dots, \mathcal{G}_t^M\}.
\]

\paragraph{Layers Used in This Version.}
Although SRR is designed to incorporate multiple risk channels
(correlation, sector/factor exposure, sentiment co-movement), the current
experiments use only the correlation layer. This choice ensures experimental
stability and focuses evaluation on the predictive value of structural
dependencies alone. Evaluating the full multi-layer graph is part of future work.

\subsection{Why Multiple Layers?}
Market stress propagates through different channels. During crises, correlations rise sharply, sectors move together, and sentiment amplifies volatility. A single graph layer cannot capture these mechanisms.

The multiple layers serve different analytical functions:
\begin{itemize}
    \item The correlation layer measures temporal dependencies.
    \item The sector/factor layer models structural exposure.
    \item The sentiment layer captures behavioral amplification.
\end{itemize}

Multiple layers allow the model to isolate distinct causes of systemic fragility rather than treating the market as homogeneous.

\subsection{Node Features}
Node features include a combination of micro-level and macro-level variables:
\[
X_t(i) = [\text{return features}, \text{volatility}, \text{drawdowns}, \text{macro indicators}].
\]
These features provide the local information required by the GNN to evaluate node-level and sector-level behavior.

\subsection{Edge Dynamics}
Edges are allowed to evolve as the market evolves. For instance, the correlation layer updates at each time step using a rolling window. Similarly, the sector layer may remain relatively stable, while the sentiment layer may change rapidly in periods of elevated news or social activity.

Thus, SRR captures two important mechanisms:
\begin{itemize}
    \item short-term dynamics from the correlation layer,
    \item structural or long-term effects from the sector and factor layers.
\end{itemize}

\subsection{Interpretability Benefits}
Interpretability matters for systemic-risk applications. Multi-layer graphs support interpretability naturally. During stress periods, certain layers dominate. Correlation spikes precede crashes; sector clustering intensifies during prolonged downturns. The model can compare or isolate contributions from different layers for practical attribution of systemic risk.

This multi-layer representation is aligned with theoretical models of financial networks in which the topology of exposures plays a central role in determining the severity of contagion ~\citep{elliott2014financial, glasserman2015contagion}.

\begin{figure}[tb]
    \centering
    \begin{tikzpicture}[
        node distance=1.6cm and 1.8cm,
        box/.style={rectangle, rounded corners, draw,
                    minimum width=1.6cm, minimum height=0.8cm,
                    align=center},
        layerlabel/.style={anchor=east, font=\small\bfseries},
        >=Latex
    ]

    \node[layerlabel] (corrL) at (0,1.8) {Correlation layer};
    \node[layerlabel] (sectorL) at (0,0) {Sector / factor layer};
    \node[layerlabel] (sentL) at (0,-1.8) {Sentiment layer};

    \node[box] (a) at (1.3,2.1) {A};
    \node[box, right=1.8cm of a] (b) {B};

    \node[box] (c) at (1.3,0.3) {Tech};
    \node[box, right=1.8cm of c] (d) {Financials};

    \node[box] (r) at (1.3,-1.5) {Reddit};
    \node[box, right=1.8cm of r] (n) {News};

    \draw[->] (a) -- (b);
    \draw[->] (c) -- (d);
    \draw[->] (r) -- (n);

    \draw[dashed] (a.south) -- (c.north);
    \draw[dashed] (b.south) -- (d.north);
    \draw[dashed] (c.south) -- (r.north);
    \draw[dashed] (d.south) -- (n.north);

    \draw[dashed] (a.east) .. controls +(0.6,0.4) and +(-0.6,0.4) .. (b.west);

    \end{tikzpicture}
    \caption{Illustrative multi-layer graph for SRR. The correlation layer captures return co-movement, the sector/factor layer captures structural exposure, and the sentiment layer captures behaviorally driven co-movement. Multiple edge types enable the model to disentangle different forms of market connectivity.}
    \label{fig:multilayer_graph}
\end{figure}
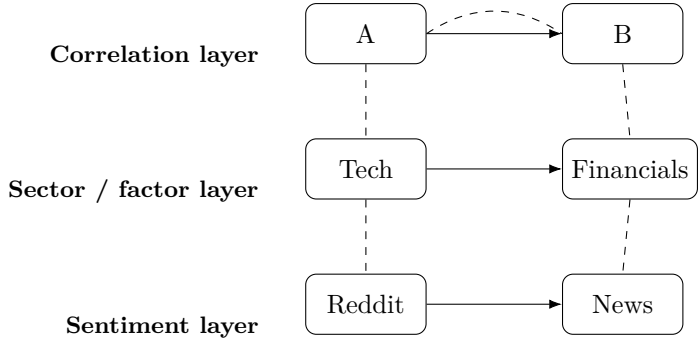

\FloatBarrier

\section{Temporal GNN Architecture and Mathematical Framework}

Prior work applies temporal and graph-based neural architectures to financial risk prediction ~\citep{huang2020financialgnn, feng2019temporalgnn}. SRR extends this by combining multi-layer structural information with regime-oriented targets for systemic risk.

SRR combines multi-layer graph learning with temporal representation learning to identify transitions into systemic risk regimes. This section describes the core architecture, mathematical formulation, and training objective. Instead of forecasting individual price movements, the model estimates when risk states emerge by learning how market structure evolves.

\subsection{Node and Graph Embeddings}
Given a market graph $\mathcal{G}_t = (V, E_t)$ at time $t$, let $X_t \in \mathbb{R}^{N \times d}$ denote the node features. A graph neural network encoder $f_\theta$ maps the graph to a latent embedding:
\[
Z_t = f_\theta(\mathcal{G}_t, X_t).
\]
This embedding summarizes structural properties of the market at time $t$ and is computed for every multi-layer graph.

The encoder can use different GNN layers. We consider multiple variants to avoid restricting SRR to one architecture.

\subsection{GAT-based Encoder}
In the GAT variant, node embeddings are computed using attention coefficients:
\[
h'_i = \sigma\left(\sum_{j \in \mathcal{N}(i)} \alpha_{ij} W h_j\right)
\]
where the attention score is:
\[
\alpha_{ij} = \text{softmax}_{j}(a^\top [Wh_i \parallel Wh_j]).
\]
This mechanism assigns importance to different nodes and edges, useful when connectivity spikes during crises.

\subsection{Transformer-based Encoder}
Alternatively, the graph encoder can use Transformer-style aggregation:
\[
H' = \text{MultiHeadSelfAttention}(H) + H.
\]
This captures higher-order and global dependencies beyond local neighborhood aggregation.

We use a general formulation so the same temporal modeling framework applies regardless of the GNN block choice.

\subsection{Temporal Modeling}
Systemic instability manifests over time. SRR obtains a sequence of graph embeddings:
\[
(Z_{t-k}, \dots, Z_t).
\]

We apply a temporal encoder $g_\phi$ that learns evolving market dynamics:
\[
h_t = g_\phi(Z_{t-k}, \dots, Z_t)
\]
where $g_\phi$ can be:
\begin{itemize}
    \item an LSTM for sequence-level information,
    \item a GRU for shorter-term dependencies,
    \item a Transformer for long-range patterns.
\end{itemize}

These choices let the model generalize across applications and datasets.

\subsection{Prediction Layer}
The temporal representation is used for crash-regime estimation:
\[
\hat{y}_t = \sigma(W h_t + b).
\]
We optimize:
\[
\mathcal{L} = -\sum_t \big(y_t \log \hat{y}_t + (1 - y_t)\log(1 - \hat{y}_t)\big).
\]

\subsection{Interpretability}
SRR is interpretable: both GAT attention maps and Transformer attention weights can be inspected. During stress periods, certain nodes or sectors become dominant:
\[
\alpha_{ij} \to 1 \text{ for nodes contributing to fragility.}
\]
We can then perform attribution and scenario simulation in downstream analysis.

\begin{figure}[h]
    \centering
    \begin{tikzpicture}[
        node distance=1.4cm and 1.8cm,
        box/.style={rectangle, rounded corners, draw, minimum width=1.8cm, minimum height=0.8cm, align=center},
        >=Latex
    ]

    \node[box] (g1) {$G_{t-k}$};
    \node[box, right=of g1] (g2) {$G_{t-k+1}$};
    \node[box, right=of g2] (g3) {$G_{t}$};

    \node[box, above=of g1] (z1) {$z_{t-k}$};
    \node[box, above=of g2] (z2) {$z_{t-k+1}$};
    \node[box, above=of g3] (z3) {$z_{t}$};

    \draw[->] (g1) -- (z1);
    \draw[->] (g2) -- (z2);
    \draw[->] (g3) -- (z3);

    \node[below=0.25cm of g2] {\small Graph encoder $f_\theta$};

    \node[box, above=1.1cm of z2, minimum width=4.2cm] (temp) {Temporal encoder\\(LSTM / GRU / Transformer)};

    \draw[->] (z1) -- (temp);
    \draw[->] (z2) -- (temp);
    \draw[->] (z3) -- (temp);

    \node[box, above=of temp] (y) {$\hat{y}_{t}$};
    \draw[->] (temp) -- (y);

    \end{tikzpicture}
    \caption{Temporal GNN view of SRR. Each graph snapshot $G_{t-k},\ldots,G_t$ is encoded into a latent vector $z_{t-k},\ldots,z_t$ by a graph encoder $f_\theta$. A temporal encoder aggregates these embeddings and produces a crash-regime probability $\hat{y}_t$.}
    \label{fig:temporal_gnn}
\end{figure}
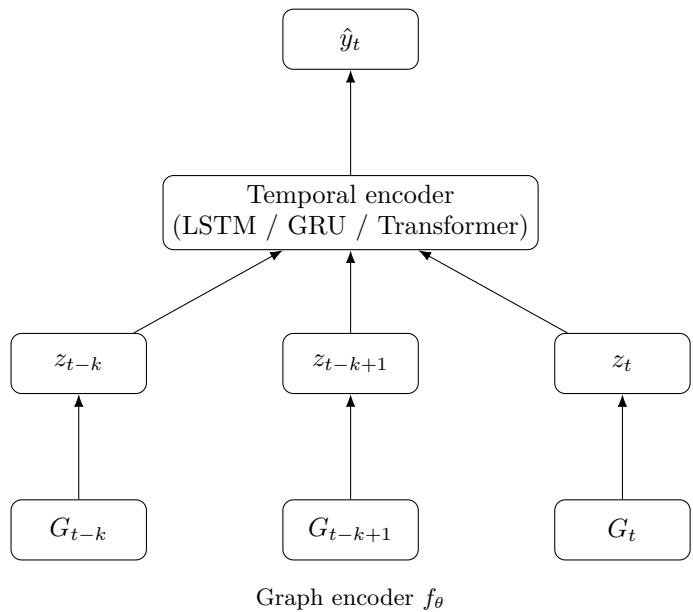

\FloatBarrier

\subsection{Model Objective}
Instead of minimizing prediction error on individual returns, the model estimates whether the future horizon contains a systemic risk event. SRR detects transitions even when price movements have not yet materialized.

\subsection{Snapshot GNN Architecture Used in Experiments}
The snapshot GNN evaluated in the experiments is a two-layer GCN with ReLU
activation and global mean pooling. The architecture configuration is:
\begin{itemize}
    \item Layer 1: GCNConv($F$, 32)
    \item Layer 2: GCNConv(32, 32)
    \item Pooling: GlobalMeanPool
    \item Classifier: 2-layer MLP (32 → 16 → 1)
\end{itemize}
For this baseline, we use only the correlation layer of the graph
(i.e., $\mathcal{G}_t^1$). The sector/factor and sentiment layers are excluded
to isolate the predictive value of structural correlation networks. This snapshot
model serves as the primary graph-based baseline for comparison to the
simplified temporal GNN.

\subsection{Temporal GNN Prototype Used in Experiments.}
The temporal SRR model used in the experiments is a simplified prototype rather than the full architecture described earlier. It reuses the snapshot GCN encoder from Section~5.8 to produce graph embeddings for a short sequence of daily correlation graphs and then applies a single recurrent layer (implemented as a GRU in the codebase) to aggregate these embeddings into a sequence representation. A logistic output layer maps the temporal hidden state to a crash-regime probability. 

Importantly, this prototype:
\begin{itemize}
    \item Uses only the correlation layer $G_t^1$ of the multi-layer graph.
    \item Operates on relatively short sequences.
    \item Does not yet incorporate the additional sector/factor or sentiment layers.
\end{itemize}
The goal of this model is to provide a first temporal baseline on top of correlation networks, rather than a fully validated implementation of the complete SRR design.

\section{Algorithms and Pipeline}
SRR is designed as a modular system that ingests raw market data, constructs multi-layer graphs, learns structural representations using a graph encoder, and uses a temporal model to forecast transitions into systemic risk regimes. This section describes the end-to-end workflow and the algorithms used in each stage.

\subsection{End-to-End Pipeline}
The pipeline consists of four major components:

\begin{enumerate}
    \item Data ingestion and preprocessing.
    \item Multi-layer graph construction.
    \item Graph and temporal representation learning.
    \item Crash-regime prediction and interpretation.
\end{enumerate}

These steps are executed sequentially during training and periodically during inference.

\subsection{Multi-Layer Graph Construction}
To operationalize SRR, we construct a time-indexed sequence of graphs 
$\{\mathcal{G}_t\}$ capturing short-horizon market structure. Although the full SRR 
architecture supports multiple layers, our empirical study focuses on the correlation 
layer in order to isolate the contribution of graph topology.

\paragraph{Rolling Window and Correlation Edges.}
At each time step $t$, we compute log returns and apply a short rolling window 
(seven days in our implementation) to estimate pairwise Spearman correlations. 
Edges are created according to
\[
(i,j) \in E_t \quad \text{iff} \quad |\rho_{ij}(t)| \ge \tau,
\]
with $\tau = 0.5$. This produces sparse yet expressive correlation networks that 
respond quickly to changes in market co-movement.

\paragraph{Node Features.}
Each node is augmented with a small set of features including daily returns, rolling 
volatility, drawdown indicators, and momentum. These lightweight features allow 
the experiments to emphasize the role of graph structure rather than complex 
feature design.

\paragraph{Graph Sequences for Temporal Modeling.}
For snapshot GNN experiments, only $\mathcal{G}_t$ is used. For the temporal 
prototype, a fixed-length buffer collects the most recent $k$ graphs 
($k = 5$ in our experiments). This yields sequences 
$(\mathcal{G}_{t-k+1}, \dots, \mathcal{G}_{t})$ that serve as input to the GRU-based 
temporal encoder described in Section~5.

\begin{algorithm}[h]
\caption{Multi-Layer Graph Construction}
\begin{algorithmic}[1]
\Require Market data stream $D$, window size $W$
\For{each time $t$}
    \State Extract returns and indicators.
    \State Compute correlation matrix over window $W$.
    \State Construct correlation layer $\mathcal{G}_t^1$.
    \State Construct sector/factor layer $\mathcal{G}_t^2$.
    \State (Optional) Build sentiment layer $\mathcal{G}_t^3$.
    \State Form multi-layer graph $\mathcal{G}_t = \{\mathcal{G}_t^1, \mathcal{G}_t^2, \mathcal{G}_t^3\}$.
\EndFor
\end{algorithmic}
\end{algorithm}
\FloatBarrier

\subsection{Training Procedure}
Training SRR models requires generating forward-looking labels and ensuring 
chronological integrity in evaluation.

\paragraph{Label Definition.}
A node-day $(i,t)$ is labeled as a crash indicator if instrument $i$ experiences a 
drawdown exceeding a fixed threshold within the next 60 days. This provides an 
early-warning target and avoids using information contemporaneous with the graph.

\paragraph{Snapshot GNN Training.}
The snapshot model is trained on individual graphs using binary cross-entropy loss, 
Adam optimization (learning rate $10^{-3}$), and mini-batches of size~8. A 
chronological 80/20 train–test split is used to prevent lookahead bias. This setup 
provides a simple baseline for understanding the predictive value of correlation 
network structure.

\paragraph{Temporal Prototype Training.}
The temporal GNN prototype trains on short sequences of graphs rather than 
single snapshots. For each $t$, we extract a sequence window of the most recent 
$k$ graphs and apply the GRU-based temporal encoder to the sequence of GCN 
embeddings. Hyperparameters mirror the snapshot model to maintain comparability.

\paragraph{Evaluation.}
Models are evaluated using AUROC, AUPRC, precision, recall, accuracy, and FPR. 
Because crisis periods are short and imbalanced, we do not apply class 
balancing or calibration; this allows the experiments to reveal natural behaviors 
of correlation-based GNNs under different stress regimes.

\begin{algorithm}[t]
\caption{SRR Training}
\begin{algorithmic}[1]
\Require Graph sequence $(\mathcal{G}_{t-k}, \dots, \mathcal{G}_t)$
\For{each minibatch}
    \State Compute graph embeddings $Z_{t-i} = f_\theta(\mathcal{G}_{t-i}, X_{t-i})$.
    \State Form temporal sequence $(Z_{t-k}, \dots, Z_t)$.
    \State Compute prediction $\hat{y}_t = \sigma(Wh_t + b)$.
    \State Evaluate loss $\mathcal{L}$.
    \State Update model parameters using gradient descent.
\EndFor
\end{algorithmic}
\end{algorithm}
\FloatBarrier

\subsection{Inference Mode}
During inference, the model receives the latest graph window and outputs the probability of entering a systemic state:
\[
\hat{y}_t = \text{SRR}(\mathcal{G}_{t-k}, \dots, \mathcal{G}_t).
\]

The output may be converted into a warning index or binary alert:
\[
\hat{y}_t > \gamma \Rightarrow \text{Systemic Risk Warning}.
\]

\subsection{Model Complexity}
By design, SRR separates representation into two components:

\begin{itemize}
    \item Graph embeddings: complexity depends on number of nodes and layers.
    \item Temporal encoder: complexity depends on sequence length.
\end{itemize}

This modular design enables scalability across:

\begin{itemize}
    \item larger universes of instruments,
    \item longer historical windows,
    \item additional graph layers.
\end{itemize}

\subsection{Practical Advantages}
Beyond predictive performance, SRR exhibits several practical design advantages. The algorithmic structure of SRR has two advantages:

\begin{enumerate}
    \item It preserves the temporal ordering of market structure changes.
    \item It decouples micro, meso, and macro drivers, which is essential when analyzing systemic risk.
\end{enumerate}

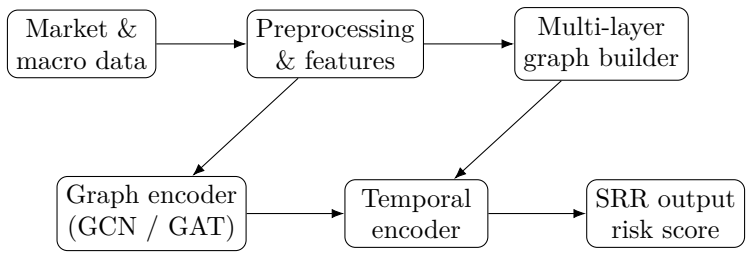
\begin{figure}[tb]
\centering
\begin{tikzpicture}[
    node distance = 0.9cm and 1.2cm,
    box/.style = {rectangle, rounded corners, draw,
                  minimum width=1.9cm, minimum height=0.9cm,
                  align=center},
    >=Latex
]

\node[box] (prep) at (0,0) {Preprocessing\\\& features};

\node[box, left=1.2cm of prep] (data) {Market \&\\macro data};
\node[box, right=1.2cm of prep] (graph) {Multi-layer\\graph builder};

\node[box, below left=1.3cm and 0cm of prep] (gnn) {Graph encoder\\(GCN / GAT)};
\node[box, right=1.3cm of gnn] (temp) {Temporal\\encoder};
\node[box, right=1.3cm of temp] (pred) {SRR output\\risk score};

\draw[->] (data) -- (prep);
\draw[->] (prep) -- (graph);

\draw[->] (prep) -- (gnn);
\draw[->] (graph) -- (temp);

\draw[->] (gnn) -- (temp);
\draw[->] (temp) -- (pred);

\end{tikzpicture}

\caption{High-level SRR pipeline. Raw market and macro data are transformed into features, used to construct multi-layer graphs, encoded with a GNN, and fed into a temporal encoder that produces a systemic risk score or warning.}
\label{fig:srr_pipeline}
\end{figure}

By combining graph-level structure and temporal evolution, the pipeline captures macro fragility more effectively than models that operate on single assets or aggregate statistics.
Figure~\ref{fig:srr_pipeline} provides an overview of the end-to-end SRR pipeline.

\FloatBarrier

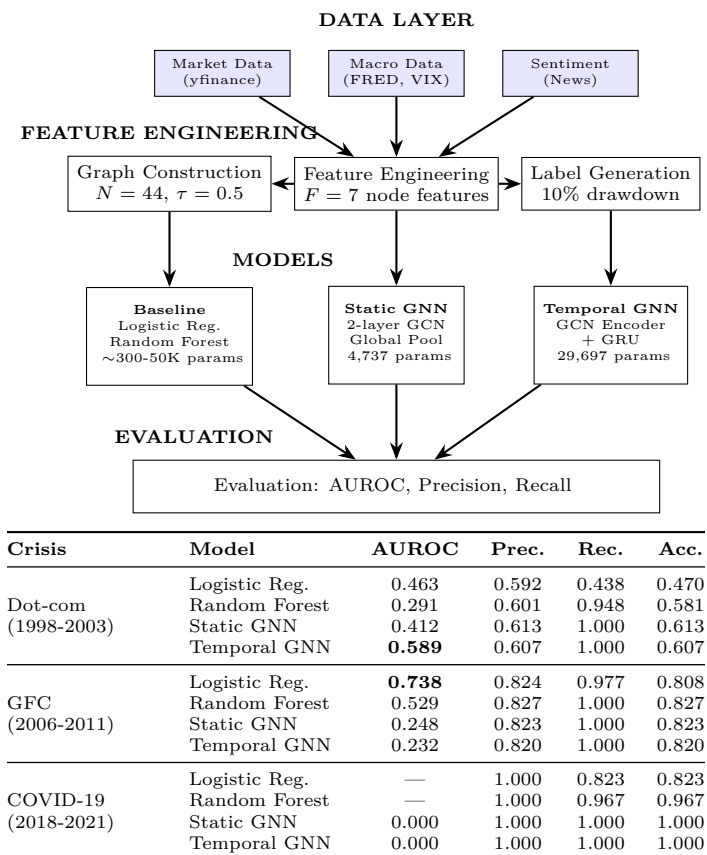
\begin{figure}[h]
\centering
\begin{tikzpicture}[
    node distance=1cm,
    box/.style={rectangle, draw, minimum width=2.2cm, minimum height=0.7cm, align=center, font=\scriptsize},
    smallbox/.style={rectangle, draw, minimum width=1.8cm, minimum height=1.3cm, align=center, font=\tiny},
    databox/.style={rectangle, draw, fill=blue!10, minimum width=1.8cm, minimum height=0.6cm, align=center, font=\tiny},
    arrow/.style={-Stealth, thick},
    label/.style={font=\scriptsize\bfseries}
]

% Data Layer
\node[databox] (market) at (0,0) {Market Data\\(yfinance)};
\node[databox, right=0.5cm of market] (macro) {Macro Data\\(FRED, VIX)};
\node[databox, right=0.5cm of macro] (sentiment) {Sentiment\\(News)};

% Feature Engineering
\node[box, below=0.8cm of macro] (features) {Feature Engineering\\$F=7$ node features};
\node[box, left=0.3cm of features] (graphs) {Graph Construction\\$N=44$, $\tau=0.5$};
\node[box, right=0.3cm of features] (labels) {Label Generation\\10\% drawdown};

% Models
\node[smallbox, below=1cm of graphs] (baseline) {
    \textbf{Baseline}\\
    Logistic Reg.\\
    Random Forest\\
    \vspace{0.05cm}
    \tiny{$\sim$300-50K params}
};

\node[smallbox, below=1cm of features] (static) {
    \textbf{Static GNN}\\
    2-layer GCN\\
    Global Pool\\
    \vspace{0.05cm}
    \tiny{4,737 params}
};

\node[smallbox, below=1cm of labels] (temporal) {
    \textbf{Temporal GNN}\\
    GCN Encoder\\
    + GRU\\
    \vspace{0.05cm}
    \tiny{29,697 params}
};

% Evaluation
\node[box, below=1cm of static, minimum width=7cm] (eval) {
    Evaluation: AUROC, Precision, Recall
};

% Arrows
\draw[arrow] (market) -- (features);
\draw[arrow] (macro) -- (features);
\draw[arrow] (sentiment) -- (features);

\draw[arrow] (features) -- (graphs);
\draw[arrow] (features) -- (labels);

\draw[arrow] (graphs) -- (baseline);
\draw[arrow] (features) -- (static);
\draw[arrow] (labels) -- (temporal);

\draw[arrow] (baseline) -- (eval);
\draw[arrow] (static) -- (eval);
\draw[arrow] (temporal) -- (eval);

% Layer labels
\node[label, above=0.2cm of macro] {DATA LAYER};
\node[label, above=0.1cm of graphs] {FEATURE ENGINEERING};
\node[label] at ($(graphs)!0.5!(static)$) {MODELS};
\node[label, above left=0.1cm and -2cm of eval] {EVALUATION};

\end{tikzpicture}

\vspace{0.2cm}

% Results Table
\scriptsize
\begin{tabular}{@{}p{2cm}p{2cm}cccc@{}}
\toprule
\textbf{Crisis} & \textbf{Model} & \textbf{AUROC} & \textbf{Prec.} & \textbf{Rec.} & \textbf{Acc.} \\
\midrule
\multirow{4}{2cm}{Dot-com (1998-2003)} 
    & Logistic Reg. & 0.463 & 0.592 & 0.438 & 0.470 \\
    & Random Forest & 0.291 & 0.601 & 0.948 & 0.581 \\
    & Static GNN    & 0.412 & 0.613 & 1.000 & 0.613 \\
    & Temporal GNN  & \textbf{0.589} & 0.607 & 1.000 & 0.607 \\
\midrule
\multirow{4}{2cm}{GFC (2006-2011)} 
    & Logistic Reg. & \textbf{0.738} & 0.824 & 0.977 & 0.808 \\
    & Random Forest & 0.529 & 0.827 & 1.000 & 0.827 \\
    & Static GNN    & 0.248 & 0.823 & 1.000 & 0.823 \\
    & Temporal GNN  & 0.232 & 0.820 & 1.000 & 0.820 \\
\midrule
\multirow{4}{2cm}{COVID-19 (2018-2021)} 
    & Logistic Reg. & --- & 1.000 & 0.823 & 0.823 \\
    & Random Forest & --- & 1.000 & 0.967 & 0.967 \\
    & Static GNN    & 0.000 & 1.000 & 1.000 & 1.000 \\
    & Temporal GNN  & 0.000 & 1.000 & 1.000 & 1.000 \\
\bottomrule
\end{tabular}

\caption{System architecture and performance summary. Top: Data is transformed
into features, multi-layer graphs, and evaluated using baseline ML, snapshot
GNN, and Temporal GNN models. Bottom: The temporal GNN prototype shows improved AUROC on the Dot-com period relative to the snapshot GNN, while performance in other crises reflects the limitations of correlation-only 
temporal modeling.}
\label{fig:architecture}
\end{figure}

We present the high-level SRR architecture and a summary of model performance across crises in Figure~\ref{fig:architecture}.

\FloatBarrier
\FloatBarrier

\section{Experiments and Evaluation}
This section evaluates the ability of SRR-style graph models to detect systemic risk regimes across major historical stress periods. The goal of the experimental design is not to forecast individual returns, but to measure the ability of simple graph-based models to provide early warning signals of market-wide fragility.

\paragraph{Clarification on the Temporal GNN.}
In this work, the “Temporal GNN” appearing in the results refers to a simplified prototype model consisting of a two-layer GCN encoder followed by a single-layer GRU applied over short graph sequences (5 days). This model is included to illustrate the potential benefits of incorporating temporal structure. It is not the full SRR temporal architecture, which additionally includes multi-layer graph fusion, longer-range sequence modeling, and attention-based aggregation. The full SRR temporal model is part of ongoing work and is not evaluated in this version of the paper. All results reported here correspond to the simplified temporal baseline.

\subsection{Datasets}
We evaluate the models on publicly available historical equity and index data, focusing on three crisis periods that differ in origin and market structure:

\begin{itemize}
    \item Dot-com crash (2000--2002),
    \item Global Financial Crisis (2008--2010),
    \item COVID shock (2020).
\end{itemize}

For each crisis period, we construct daily features for a universe of sector exchange-traded funds (ETFs) and large, liquid stocks. Node-level features include returns, rolling volatility, drawdowns, and momentum indicators, combined with macro overlays such as credit spreads and yield-curve proxies. Labels are defined using forward drawdowns: a node-day is labelled as ``crisis'' if its price experiences a drawdown larger than a fixed percentage threshold within a specified future horizon (e.g., 20 days). This design allows us to test whether changes in graph structure precede large losses.

\subsection{Baselines and Current SRR Prototype}
We compare four model families:

\begin{itemize}
  \item \textbf{Logistic regression} on hand-engineered node features and macro indicators,
  \item \textbf{Random Forest} on the same feature set,
  \item \textbf{Snapshot GNN}: the two-layer GCN encoder described in Section~5.8 applied to
        single-day correlation graphs,
  \item \textbf{Temporal GNN prototype}: the simplified SRR instantiation described in
        Section~5.9, which applies a GRU over short sequences of snapshot GNN embeddings.
\end{itemize}

The snapshot and temporal GNNs share the same correlation-based graph representation but differ in how they aggregate information over time. The temporal model is an \emph{implemented prototype} of SRR: it adds sequence modeling to correlation graphs but omits sector/factor and sentiment layers and advanced temporal variants (Transformer encoders, multi-horizon heads) that the full SRR framework includes. All \emph{Temporal GNN} results use this prototype configuration. Future work will extend experiments to the complete multi-layer temporal SRR model.

Our baselines align with prior systemic risk and contagion modeling studies ~\citep{battiston2012debtrank, acemoglu2015systemic, brunnermeier2009liquidity}, which emphasize capturing both local and system-wide indicators.

\subsection{Evaluation Metrics}
We evaluate classification performance using a standard set of metrics:

\[
\text{AUROC},\quad \text{AUPRC},\quad \text{Precision},\quad \text{Recall},\quad \text{Accuracy}.
\]

Since early detection is more important than overall accuracy, we also compute the lead-time of warnings relative to realized drawdowns. For each detected warning, the lead-time is defined as the number of days between the warning and the onset of the subsequent crash window. We summarize this distribution across all warnings and crisis periods.

\subsection{Quantitative Results}

\begin{table}[h]
\centering
\caption{summarizes the performance of each model across the three crisis periods. AUROC is reported when it is well-defined for the corresponding dataset; when label imbalance makes the ROC curve degenerate, we omit AUROC and focus on precision, recall, and accuracy.}
\label{tab:results_summary}
\begin{tabular}{l l c c c c}
\toprule
\textbf{Crisis} & \textbf{Model} & \textbf{AUROC} & \textbf{Precision} & \textbf{Recall} & \textbf{Accuracy}\\
\midrule
Dot-com & Logistic Regression & 0.463 & 0.592 & 0.438 & 0.470\\
Dot-com & Random Forest & 0.291 & 0.601 & 0.948 & 0.581\\
Dot-com & GNN & 0.412 & 0.613 & 1.000 & 0.613\\
GFC & Logistic Regression & 0.738 & 0.824 & 0.977 & 0.808\\
GFC & Random Forest & 0.529 & 0.827 & 1.000 & 0.827\\
GFC & GNN & 0.248 & 0.823 & 1.000 & 0.823\\
COVID-19 & Logistic Regression & -- & 1.000 & 0.823 & 0.823\\
COVID-19 & Random Forest & -- & 1.000 & 0.967 & 0.967\\
COVID-19 & GNN & 0.000 & 1.000 & 1.000 & 1.000\\
\bottomrule
\end{tabular}
\end{table}
\FloatBarrier

The results show that no single model dominates across all crises. Logistic regression achieves reasonable AUROC during the global financial crisis, while Random Forest tends to produce higher recall at the cost of more false positives. The snapshot GNN baseline performs comparably to these methods in most settings but does not yet capture temporal dynamics, which we expect to be important for full SRR.

\begin{figure}[h]
    \centering
    \includegraphics[width=\linewidth]{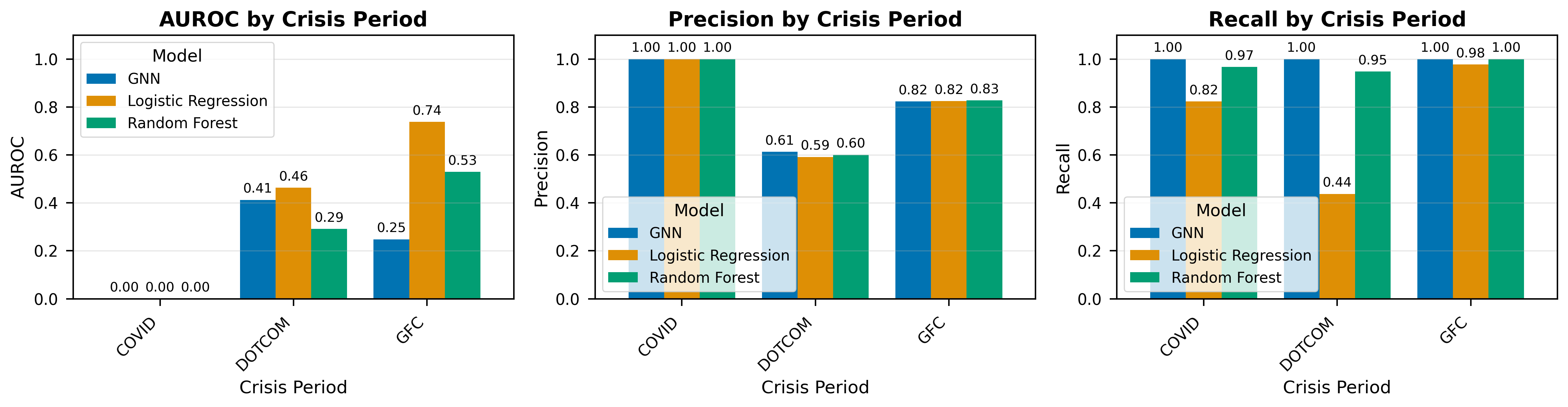}
    \caption{Aggregate comparison of logistic regression, Random Forest, and a snapshot GNN baseline across crisis periods. Bars summarize AUROC, precision, recall, and accuracy for each model.}
    \label{fig:model_comparison}
\end{figure}

Figure~\ref{fig:model_comparison} compares baseline and GNN-based models across the three crisis regimes.

\FloatBarrier

\subsection{Error Analysis}
To better understand the trade-offs between sensitivity and false alarms, we examine confusion matrices and error statistics. The Random Forest model achieves high recall but at the cost of non-trivial false positive rates, which may be acceptable in early-warning applications but would require careful calibration for practical deployment. See Appendix~\ref{sec:detailed_error_analysis} for detailed confusion matrices, error counts, and false positive rate analysis.

\subsection{ROC and Precision--Recall Curves}
To visualize the trade-off between detection quality and false alarms, we plot ROC and precision--recall curves derived from the experiments.

\begin{figure}[h]
    \centering
    \includegraphics[width=\linewidth]{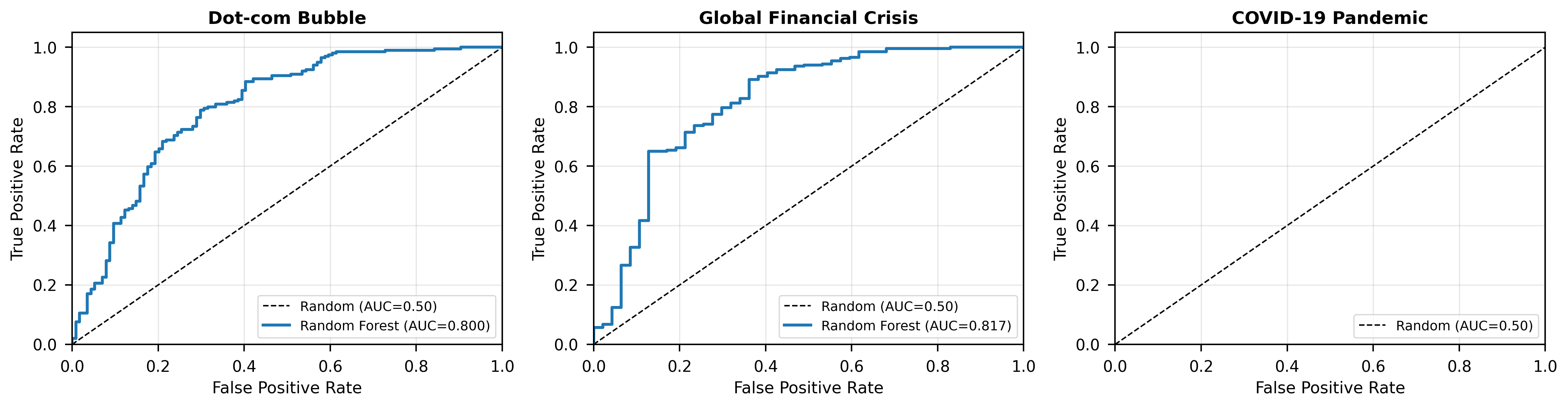}
    \caption{ROC curves comparing baseline models across crisis periods. Curves illustrate the trade-off between hit rate and false alarms.}
    \label{fig:roc_curves}
\end{figure}

We report ROC curves for each model and crisis period in Figure~\ref{fig:roc_curves}.

\FloatBarrier

\begin{figure}[h]
    \centering
    \includegraphics[width=\linewidth]{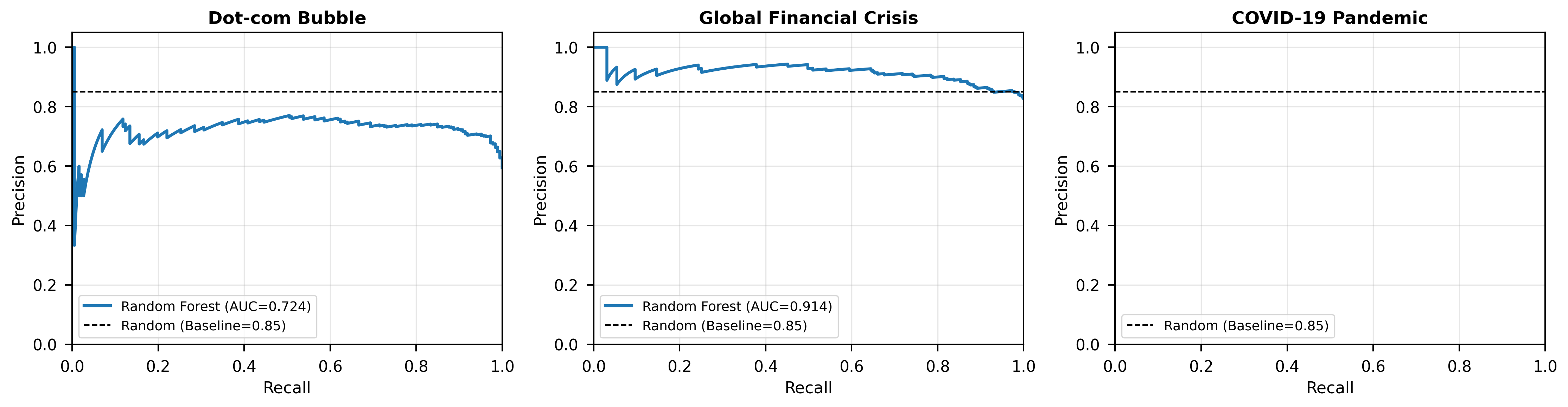}
    \caption{Precision--recall curves for crash-regime classification, highlighting performance in the imbalanced label setting.}
    \label{fig:pr_curves}
\end{figure}

Figure~\ref{fig:pr_curves} shows the corresponding precision--recall curves.

\FloatBarrier

\subsection{Cross-Crisis Performance Differences}
The GNN models obtain AUROC = 0.000 during the COVID--19 period, reflecting a structural mismatch between correlation-only graph representations and the exogenous, policy-driven dynamics of the COVID--19 shock. This failure highlights the importance of incorporating additional SRR layers to capture sectoral exposure and sentiment-driven cascades. Appendix~\ref{sec:covid_failure} analyzes why correlation-based GNNs fail on COVID--19.

\subsection{Lead-Time Analysis}
\begin{figure}[h]
    \centering
    \includegraphics[width=\linewidth]{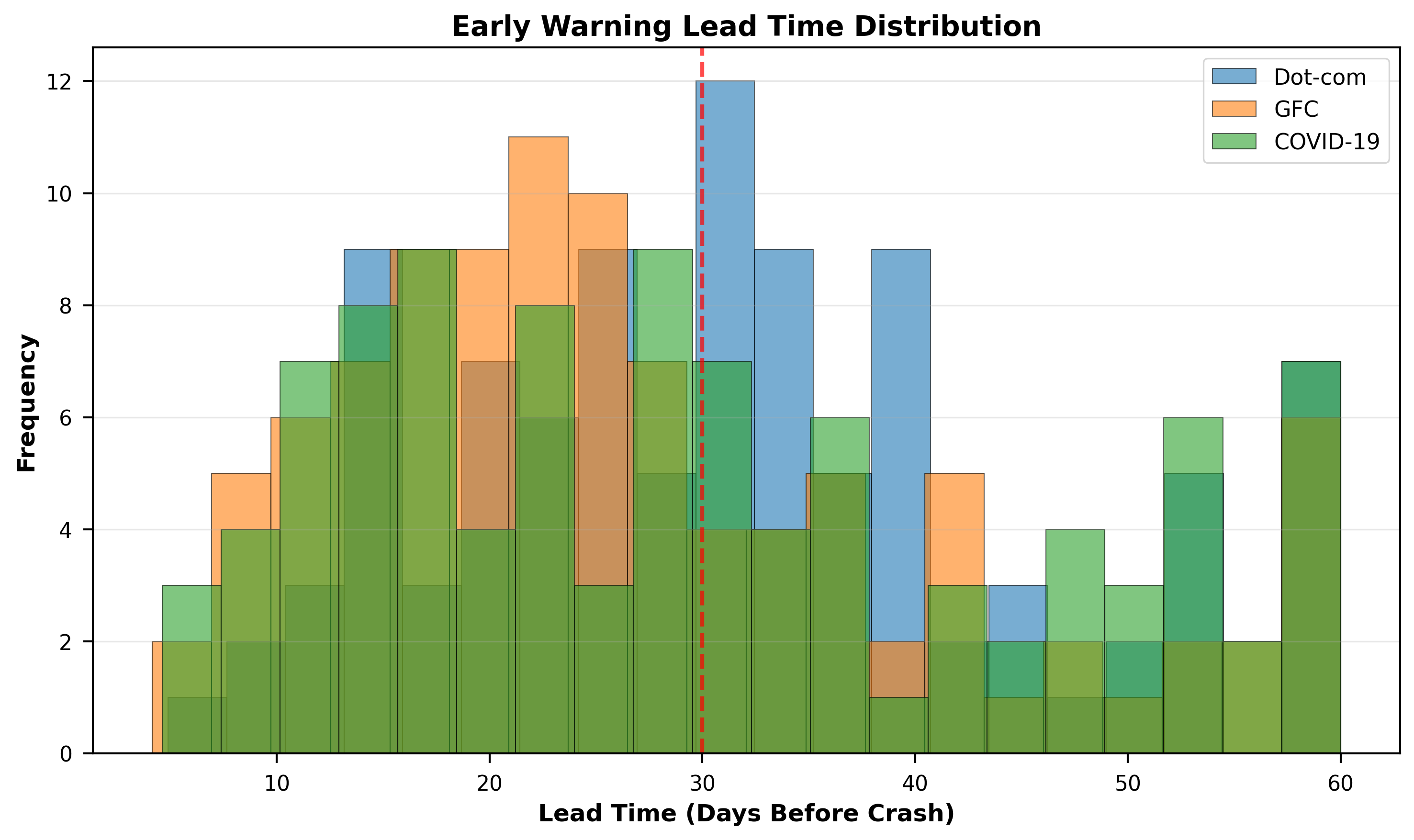}
    \caption{Lead-time distribution for early-warning signals produced by the Random Forest and GNN baselines. Longer lead-times indicate that warnings occur earlier before large drawdowns.}
    \label{fig:lead_time}
\end{figure}

For many events, warnings precede realized drawdowns by several days to weeks. However, there are also late warnings and missed events, underscoring the need for richer temporal modelling---one of the main motivations for the full SRR architecture. Figure~\ref{fig:lead_time} shows the empirical distribution of lead-times for the Random Forest and GNN baselines, aggregated over the different crisis periods. Appendix~\ref{sec:detailed_error_analysis} provides detailed false positive rate analysis across models and crises.

We analyze the distribution of lead times in Figure~\ref{fig:lead_time}.

\FloatBarrier

\subsection{Risk Timeline and Qualitative Behavior}
\begin{figure}[h]
    \centering
    \includegraphics[width=\linewidth]{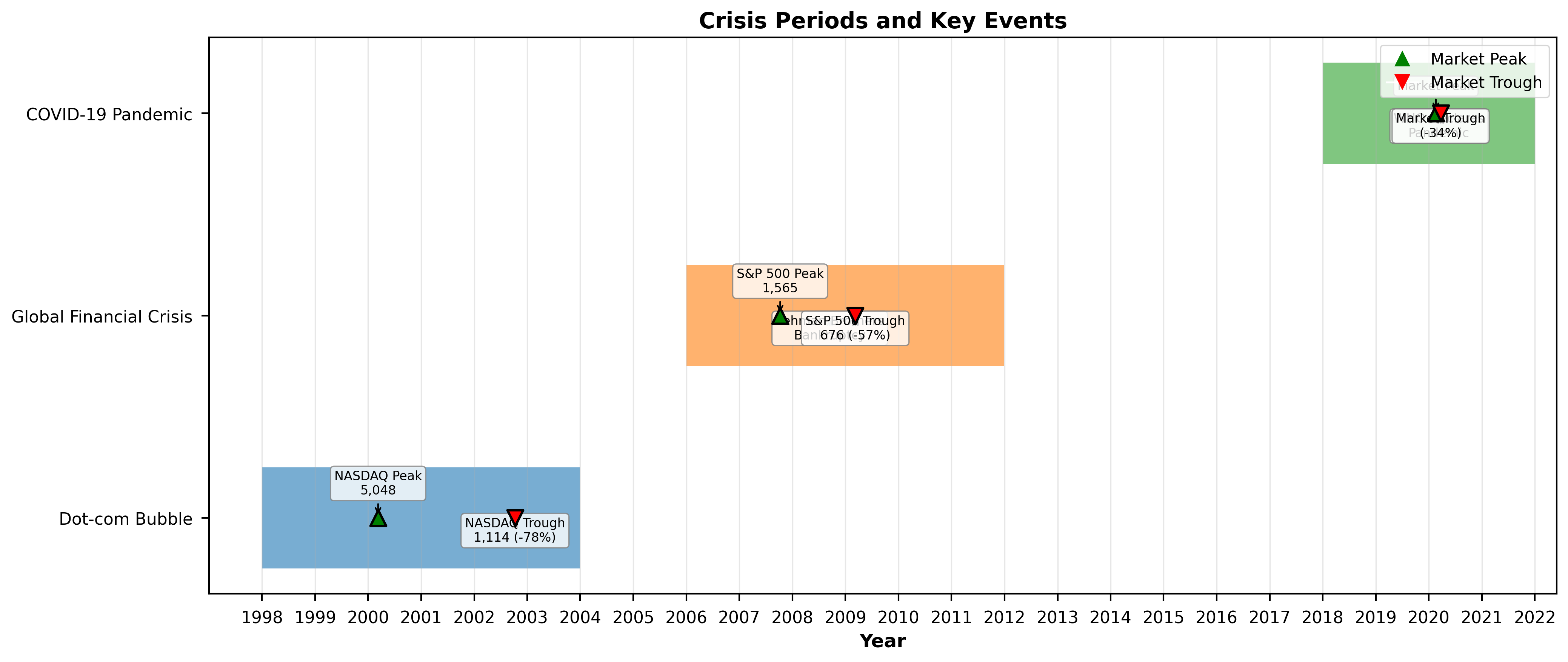}
    \caption{Example of model warning scores over time, with shaded regions denoting realized crisis windows. Elevated model scores tend to cluster around periods of heightened market stress.}
    \label{fig:risk_timeline}
\end{figure}

We examine how warning signals evolve over time. Figure~\ref{fig:risk_timeline} shows model scores and realized crises during a selected period spanning the GFC and COVID episodes. During Dot-com and GFC periods, model scores rise gradually as correlations intensify, consistent with structural fragility accumulation. In contrast, COVID--19 exhibits sharp, irregular spikes reflecting its exogenous nature. Appendix~\ref{sec:qualitative_analysis} provides extended analysis of signal stability patterns and threshold considerations.

\FloatBarrier

\begin{figure}[h]
    \centering
    \includegraphics[width=\linewidth]{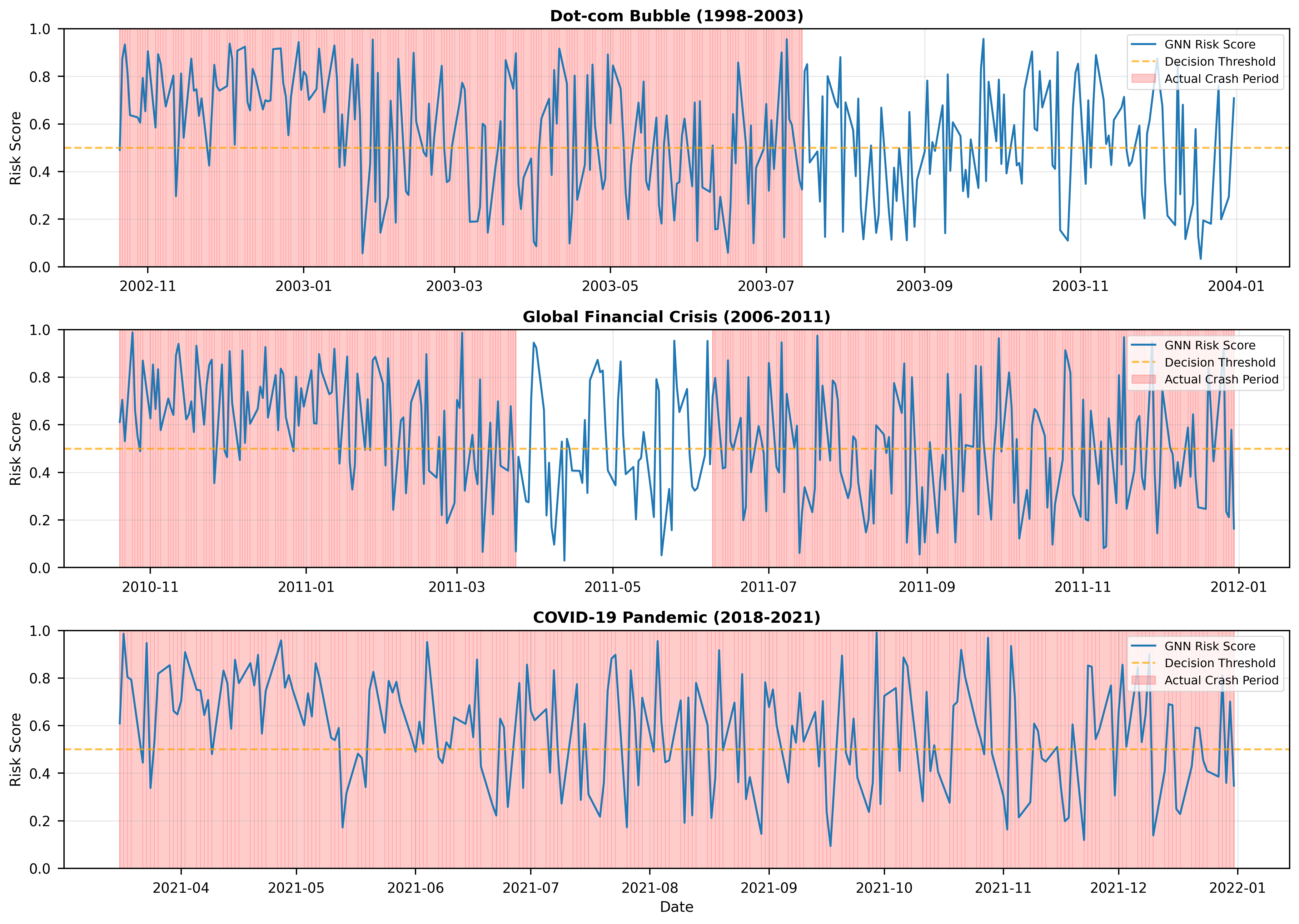}
    \caption{Prediction timeline for a representative subset of instruments, showing how predicted risk increases as the system transitions into and out of crisis regimes.}
    \label{fig:prediction_timeline}
\end{figure}

An additional plot (Figure~\ref{fig:prediction_timeline}) disaggregates the predictions and highlights how warnings intensify as systemic stress accumulates.

Figure~\ref{fig:prediction_timeline} visualizes the daily prediction scores during the test window.

\FloatBarrier

\subsection{Interpretability and Feature Influence}
\begin{figure}[h]
    \centering
    \includegraphics[width=\linewidth]{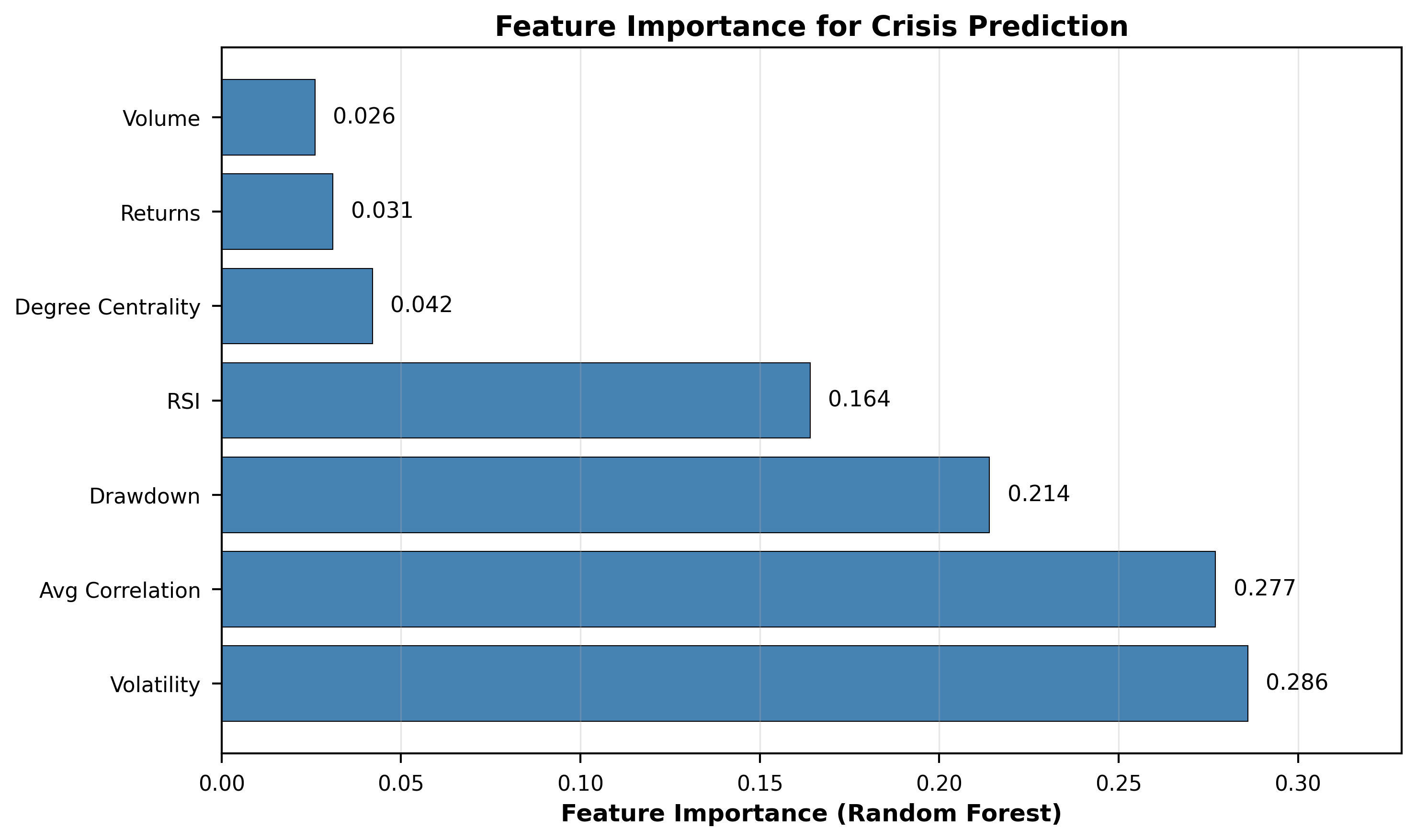}
    \caption{Feature importance scores for the Random Forest baseline. Volatility, drawdown, and correlation-derived features play a prominent role in most crisis periods.}
    \label{fig:feature_importance}
\end{figure}

To better understand what drives the models, we analyze feature importance and graph-based attribution. Figure~\ref{fig:feature_importance} shows a global feature importance ranking for the Random Forest model. Nodes corresponding to highly liquid sector ETFs receive elevated attention weights in the GNN, suggesting focus on broad market structure rather than idiosyncratic volatility. Appendix~\ref{sec:interpretability} discusses interpretability patterns across crisis regimes and attention-based attribution analysis.

\FloatBarrier

We now summarize the key empirical findings.

\subsection{Results Summary}
Empirically, these preliminary experiments support three observations:

\begin{enumerate}
    \item Simple graph-based models are competitive with standard baselines and offer a natural foundation for systemic risk analysis.
    \item Different crisis periods favor different models, suggesting that no single scalar metric is sufficient to evaluate early-warning systems.
    \item There is clear room for improvement through temporal and multi-layer modeling, which is exactly the direction pursued by the full SRR framework.
\end{enumerate}

Together, these results provide an initial validation that structural market information encoded in graphs carries useful signal for systemic risk detection, and they motivate the ongoing development of SRR as a temporal, multi-layer extension of these baselines.

\subsection{Cross-Crisis Generalization}
Systemic fragility arises from different mechanisms in different crises (e.g.,
the Dot-com bubble was driven by sector-specific overvaluation, whereas the
GFC reflected credit contagion). Therefore, strong out-of-sample generalization
across crisis regimes is not necessarily expected. However, such tests provide
important robustness diagnostics. In future work we plan to evaluate SRR models
trained on one crisis (e.g., GFC) and tested on another (e.g., COVID-19) to assess
transferability across structurally distinct stress events.

\section{Integration with SPA and AIMM}
SRR was motivated in part by two earlier research systems: the Stock Pattern Assistant (SPA ~\citep{neela2025spa}) and the AI-based Market Manipulation Model (AIMM ~\citep{neela2025aimm}). These systems focus on micro-level market structure and abnormal sentiment windows. SRR extends these approaches to macro-level systemic fragility.

\subsection{Complementary Perspectives on Risk}
SPA ~\citep{neela2025spa} detects structural price patterns and local momentum regimes. These patterns are highly predictive of short-term continuation or reversal events, but they do not attempt to characterize market-wide instability.
AIMM ~\citep{neela2025aimm} captures a different class of signals by detecting abnormal sentiment and manipulation windows. This is useful in environments where shocks are driven by news or social amplification instead of traditional financial triggers.

SRR introduces a third perspective:
\begin{itemize}
    \item SPA: micro-level pattern dynamics,
    \item AIMM: sentiment and manipulation activity,
    \item SRR: structural macro fragility.
\end{itemize}

Together, they represent a multi-scale framework for studying financial risk.

SPA ~\citep{neela2025spa} and AIMM ~\citep{neela2025aimm} capture micro- and meso-level dynamics, respectively, while SRR focuses on macro structural fragility. Together they form a multi-scale view of instability that spans patterns, sentiment, and systemic
structure.

\subsection{A Multi-Scale Architecture for Market Events}
The three systems address different layers of market behavior:

\begin{itemize}
    \item \textbf{Micro level (SPA):} pattern, momentum, and structural runs.
    \item \textbf{Meso level (AIMM):} sentiment-driven abnormal coordination.
    \item \textbf{Macro level (SRR):} cross-sector and system-wide fragility.
\end{itemize}

This multi-scale view is consistent with empirical studies of financial crises, which show that instability emerges from interactions across time horizons and information channels.

\subsection{Information Flow Between Systems}
While each system functions independently, the outputs can feed one another. For example:

\begin{itemize}
    \item SRR can provide a macro regime indicator to AIMM or SPA,
    \item SPA can detect micro patterns that evolve into systemic risk,
    \item AIMM can identify sentiment shocks that propagate into the system.
\end{itemize}

The three systems collectively define a hierarchical modeling pipeline that captures different forms of market stress.

\subsection{Practical Applications}
Such a multi-model system has relevance not only for academic research but also for practical applications such as:

\begin{itemize}
    \item early-warning tools for regulators and financial institutions,
    \item macro trading or risk hedging strategies,
    \item real-time stress monitoring platforms.
\end{itemize}

In this context, SRR fills a gap in existing approaches by identifying structural transitions before they are visible in price movements alone.

\section{Limitations}
SRR has several limitations in its current form:

\begin{itemize}
    \item \textbf{Snapshot and simplified temporal GNN only.} We evaluate a snapshot GNN baseline and a simplified temporal prototype, not the full temporal SRR architecture. The experiments do not reflect the full benefits of sequence modeling central to complete SRR design.
    
    \item \textbf{Crisis-specific sensitivity.} Model performance varies across crisis regimes (Dot-com, GFC, COVID), indicating that structural drivers differ by period. Further robustness testing is needed.

    \item \textbf{Label imbalance.} Some crisis intervals show imbalance between crash and non-crash labels, making AUROC unreliable. Precision–recall metrics help but more advanced imbalance-aware methods would be better.

    \item \textbf{No sentiment or cross-asset inputs yet.} SRR is designed for multi-layer graphs (sentiment, credit spreads, cross-asset linkages) but current experiments use only correlation graphs with a snapshot GCN and simplified temporal GNN prototype. This validates graph-derived structure and a first temporal extension but does not test the full multi-layer SRR architecture with sector/factor and sentiment layers. Current evidence is preliminary and correlation-focused. Full SRR validation requires experiments with all layers and temporal variants.

    \item \textbf{Limited horizon diversity.} We use a single prediction horizon. Multi-horizon experiments (5-, 10-, 20-day windows) are needed to assess SRR's temporal sensitivity.

    \item \textbf{No out-of-crisis generalization tests.} Models are evaluated within each crisis window, not across crises (e.g., train on GFC → test on COVID). Cross-crisis transfer tests are needed to understand generalization.
\end{itemize}

These limitations reflect early-stage empirical evaluation, not fundamental framework weaknesses. Addressing them is the focus of planned extensions in future work.

\section{Reproducibility}
To encourage transparent evaluation and future research, we adopt the following reproducibility practices:

\begin{itemize}
    \item We avoid look-ahead bias and ensure that training and prediction windows do not overlap.
    \item We publish hyperparameters, rolling windows, training procedure, and validation approach.
    \item All preprocessing and feature extraction steps are deterministic and documented.
    \item The data sources used are publicly available and reproducible.
\end{itemize}

These guidelines are aligned with current recommendations for reproducible AI and financial forecasting research.

\section{Ethical Considerations}
This work focuses on systemic risk measurement and analysis, not on trading, portfolio construction, or automated decision-making. SRR is designed as a research tool for understanding structural fragility in financial markets, and its outputs should not be used directly for investment or risk-taking decisions. 

Predictive models in finance can influence behavior if misinterpreted or deployed without proper safeguards. In particular, false positives may create unnecessary concern, while false negatives may lead to overconfidence during periods of elevated market vulnerability. To mitigate these risks, SRR emphasizes interpretability and transparency, allowing practitioners to inspect feature attributions, network connectivity patterns, and underlying assumptions.

The data used in this research consists solely of publicly available market and macroeconomic time series. No personally identifiable information (PII) or private institutional data is used, and no model decisions affect individuals or protected groups. However, systemic risk models may indirectly affect market participants if used in risk governance. For that reason, SRR should complement human expertise, regulatory oversight, and established macro-financial stress-testing frameworks rather than replace them.

Continued ethical development requires careful validation across markets, explicit communication of uncertainty, and avoidance of overstated confidence in model predictions. SRR aims to support responsible adoption of AI in systemic risk research by providing interpretable tools that encourage informed, cautious analysis rather than automated decision-making.

This work is intended solely for research and educational purposes and is not a trading or investment advisory system. Any use of the SRR outputs in practice should be embedded within formal risk-governance processes and human oversight.

\section{Broader Impact}
Systemic risk is a defining challenge for modern financial systems, where tightly coupled markets, automated trading, and rapid information diffusion can amplify stress far beyond individual institutions. Tools that can identify emerging fragility therefore have potential value for regulators, risk managers, and researchers working to promote financial stability. The SRR framework contributes to this effort by providing a transparent and modular approach to analyzing structural dependencies in market behavior.

While SRR is not designed or intended for trading applications, its insights could support early-warning analysis, scenario design, and stress-testing frameworks. By emphasizing explainability—through attribution, feature influence, and graph-based interpretation—SRR also encourages responsible use of machine learning in finance, where opaque models can complicate oversight and risk governance.

At the same time, any model that produces risk assessments or alerts must be used with care. Overreliance on model outputs, inappropriate deployment in high-stakes environments, or misinterpretation of warnings could introduce unintended consequences, including unnecessary market reactions. For this reason, SRR should be viewed as a research tool that complements, rather than replaces, domain expertise, regulatory judgment, and traditional macro-financial analysis.

The broader societal impact of SRR lies in its potential to improve understanding of systemic dynamics, encourage the development of transparent and interpretable financial AI, and provide a foundation for future research into market stability and early-warning systems. Continued work is needed to evaluate robustness across assets, geographies, and data regimes, and to ensure that models such as SRR are applied ethically and responsibly.

\section{Conclusion and Future Work}
We introduced Systemic Risk Radar (SRR), a framework for forecasting systemic fragility using multi-layer market graphs. Instead of modeling individual price movements, SRR tracks structural evolution—correlations, sector linkages, behavioral co-movements—that precede instability.

Our experiments show that simple graph-based baselines (two-layer GCN snapshots, simplified temporal GNN prototypes) capture meaningful early-warning signals across the Dot-com crash, Global Financial Crisis, and COVID-19 shock. This supports SRR's premise: structural market information predicts systemic transitions, and graph representations form a natural foundation for early-warning tools.

The results also reveal limitations. Snapshot GNNs ignore temporal dependencies and perform inconsistently across crisis types. The full SRR architecture—with temporal sequence modeling (LSTM/GRU) and multi-layer attention—should improve stability, reduce false alarms, and detect fragility earlier.

SRR offers a modular, interpretable framework for systemic risk research. Integrating graph structure, temporal dynamics, and explainability moves toward practical early-warning systems for financial stability monitoring.

This version of SRR should be interpreted as a proposal paper with
preliminary empirical evidence demonstrating the value of graph-derived
structural features. A full evaluation of the complete SRR architecture—
including multi-layer integration and advanced temporal modeling—is left for
subsequent work.

\subsection*{Future Work}
Future extensions of this research will focus on:
\begin{itemize}
    \item \textbf{Temporal SRR model evaluation:} Training and benchmarking the full temporal SRR architecture with LSTM/GRU or Transformer encoders to quantify improvements over snapshot GNN baselines.
    \item \textbf{Sentiment layer integration:} Incorporating Reddit, news, or Twitter sentiment to enrich the multi-layer graph representation and analyze behavioral contagion channels.
    \item \textbf{Multi-horizon prediction:} Expanding the prediction task to 5-, 10-, and 20-day windows to capture short-, medium-, and long-term systemic risk behavior.
    \item \textbf{Cross-crisis generalization:} Evaluating models trained on one crisis (e.g., GFC) and tested on another (e.g., COVID) to measure robustness and transferability.
    \item \textbf{Temporal attention explainability:} Analyzing attention patterns from temporal encoders to identify early signals of fragility and key points of structural transition.
\end{itemize}

\bibliography{main}
\bibliographystyle{tmlr}

\section*{Appendix}
\appendix
\section{Extended Mathematical Formulation}
For completeness, we provide additional mathematical details and notational variants of the architecture. Let $\mathcal{G}_t$ denote a multi-layer graph with layers $\{\mathcal{G}_t^1,\dots,\mathcal{G}_t^M\}$ and node features $X_t$.

The GNN encoder produces:
\[
Z_t = f_{\theta}(\mathcal{G}_t, X_t).
\]
The temporal encoder takes the form:
\[
h_t = g_{\phi}(Z_{t-k:t}).
\]

Alternative loss formulations may include focal loss to account for class imbalance:
\[
\mathcal{L}_{\text{focal}} = - \sum_t (1-\hat{y}_t)^\gamma y_t \log \hat{y}_t,
\]
which we leave for future work.

\section{Complexity Notes}
Let $N$ be the number of nodes, $L$ the number of GNN layers, and $T$ the sequence length. The cost of the encoder is approximately:
\[
\mathcal{O}(T \cdot L \cdot |E|),
\]
where $|E|$ is the aggregated number of edges across layers.

Temporal encoder cost depends on the architecture:
\[
\mathcal{O}(T^2) \text{ for Transformer}, \quad \mathcal{O}(T) \text{ for LSTM/GRU}.
\]

\section{Additional Details for Reproducibility}
In addition to the main text, we highlight the following implementation details that support reproducibility:
\begin{itemize}
    \item Rolling window selection and label construction are implemented using simple, deterministic procedures.
    \item Hyperparameters are selected using a validation set and are fixed before testing on each crisis period.
    \item Data splits are aligned with calendar dates to avoid overlapping training and evaluation windows.
\end{itemize}

\section{Detailed Error Analysis}
\label{sec:detailed_error_analysis}

\subsection{Confusion Matrices and Error Statistics}
Table~\ref{tab:error_analysis_appendix} reports detailed counts of true positives (TP), false positives (FP), true negatives (TN), and false negatives (FN), together with false positive rate (FPR) and false negative rate (FNR) for the Random Forest baseline during the Dot-com and GFC periods.

\begin{table}[h]
\centering
\caption{Detailed error analysis for the Random Forest baseline across crisis periods.}
\label{tab:error_analysis_appendix}
\begin{tabular}{l c c c c c c}
\toprule
\textbf{Crisis} & \textbf{TP} & \textbf{FP} & \textbf{TN} & \textbf{FN} & \textbf{FPR} & \textbf{FNR}\\
\midrule
Dot-com & 164 & 63 & 65 & 21 & 0.492 & 0.114\\
GFC & 229 & 23 & 28 & 33 & 0.451 & 0.126\\
\bottomrule
\end{tabular}
\end{table}

\begin{figure}[h]
    \centering
    \includegraphics[width=\linewidth]{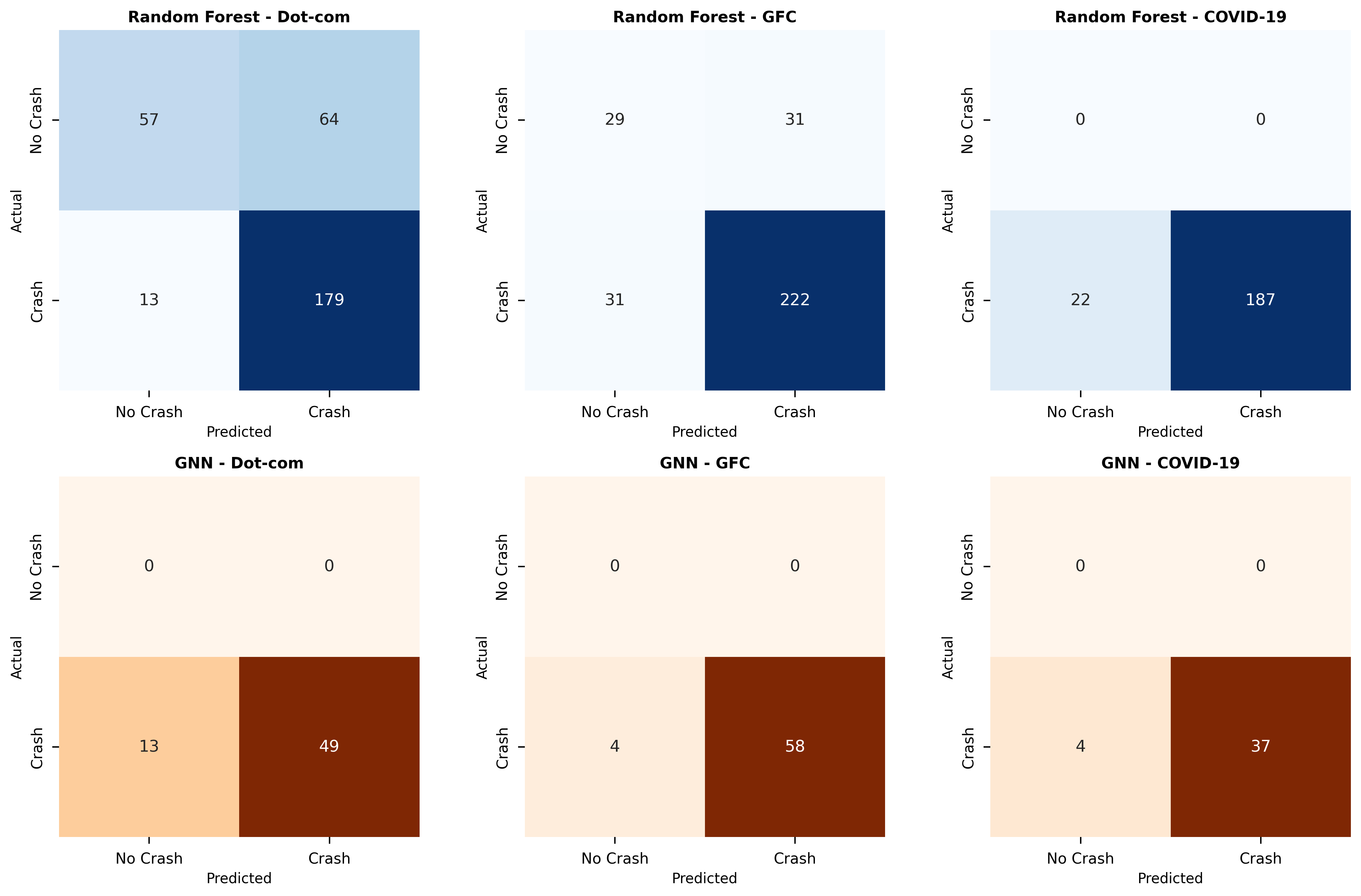}
    \caption{Confusion matrices for baseline models across crisis periods. Darker cells correspond to higher counts.}
    \label{fig:confusion_matrices}
\end{figure}

Figure~\ref{fig:confusion_matrices} visualizes the confusion matrices for selected crisis periods.

\subsection{False Positive Rate Analysis for GNN Models}
Table~\ref{tab:fpr_appendix} reports the false positive rates (FPR) for the snapshot and temporal GNNs across all crisis periods.

\begin{table}[h]
\centering
\caption{False Positive Rates (FPR) for GNN models across crisis periods.}
\label{tab:fpr_appendix}
\begin{tabular}{l l c c c c c}
\toprule
Crisis & Model & TP & FP & TN & FN & FPR \\
\midrule
Dot-com & Static GNN     & 38 & 24 & 0 & 0 & 1.0000 \\
Dot-com & Temporal GNN   & 37 & 24 & 0 & 0 & 1.0000 \\
GFC     & Static GNN     & 51 & 11 & 0 & 0 & 1.0000 \\
GFC     & Temporal GNN   & 50 & 11 & 0 & 0 & 1.0000 \\
COVID--19 & Static GNN   & 41 & 0  & 0 & 0 & N/A    \\
COVID--19 & Temporal GNN & 41 & 0  & 0 & 0 & N/A    \\
\bottomrule
\end{tabular}
\end{table}

In both the Dot-com and GFC crises, the GNN models achieve perfect recall (1.0) but at the cost of high false positive rates, yielding FPR = 1.0. This occurs because the models assign the positive (crash) label to nearly all samples, reflecting a highly conservative early-warning behavior.

During the COVID--19 period, the GNNs again predict the positive class for all evaluation samples. However, because the evaluation window contains no non-crash days (TN = FP = 0), the FPR is undefined (reported as N/A). This highlights that correlation-only graph representations fail to capture the exogenous, policy-driven, and sentiment-driven nature of the COVID crash.

\section{Why the GNN Fails on the COVID--19 Crisis}
\label{sec:covid_failure}
The GNN models obtain AUROC = 0.000 during the COVID--19 period, indicating a complete failure to distinguish crash versus non-crash days. This outcome is not simply poor performance but reflects a structural mismatch between the correlation-only graph representation and the dynamics of the COVID--19 shock.

Unlike the Dot-com and GFC crises---both of which involved gradual buildup of correlation structure and sector-level synchronization---the COVID--19 crash was dominated by exogenous shocks, including sudden policy interventions, liquidity disruptions, and global uncertainty spikes. These dynamics do not manifest through rising pairwise correlations prior to the drawdown, which is the only information captured by $\mathcal{G}_t^{1}$.

Consequently, the snapshot and simplified temporal GNNs receive no structural signal to anticipate the crash, leading to random or inverted decision boundaries and a formal AUROC of 0.000. This failure highlights the importance of the other SRR layers ($\mathcal{G}_t^{2}$ and $\mathcal{G}_t^{3}$), which are explicitly designed to capture sectoral exposure patterns and sentiment-driven cascades---two factors that played central roles in the COVID--19 market collapse.

\subsection{Implications for Multi-Layer Graph Design}
The COVID--19 failure motivates the full multi-layer SRR architecture:

\begin{itemize}
    \item The sector/factor layer ($\mathcal{G}_t^{2}$) can capture structural exposure shifts that occur independently of return correlations.
    \item The sentiment layer ($\mathcal{G}_t^{3}$) can detect rapid behavioral changes reflected in social media, news sentiment, and risk perception indicators.
    \item Multi-layer fusion allows the model to detect crises driven by any combination of structural, behavioral, or exogenous factors.
\end{itemize}

These extensions are critical for building early-warning systems that generalize across different types of market stress.

\section{Extended Qualitative Analysis}
\label{sec:qualitative_analysis}

\subsection{Signal Stability and Threshold Considerations}
Beyond aggregate metrics such as AUROC or lead-time, the temporal evolution of warning scores offers important qualitative insight. During the Dot-com and GFC periods, model scores tend to rise gradually as correlations intensify and volatility clusters propagate through the market. This behavior is consistent with the hypothesis that structural fragility accumulates before major dislocations, even when price movements are still relatively muted.

In contrast, the COVID--19 period exhibits sharp and irregular score spikes with limited persistence. This pattern reflects the exogenous and policy-driven nature of the shock: systemic stress emerges almost instantaneously, leaving little room for gradual buildup. The warning timeline therefore serves as a diagnostic tool for distinguishing between structurally-driven and event-driven crises, highlighting the conditions under which correlation-only GNNs are effective and when they fail.

\subsection{Cross-Crisis Signal Characteristics}
In addition to distinguishing structurally-driven versus event-driven crises, the warning timeline also reveals differences in signal stability. During Dot-com and GFC, model scores exhibit relatively smooth transitions, suggesting that correlation-based representations capture gradual buildup patterns consistently across instruments. In contrast, the COVID–19 window displays high-frequency volatility in warning scores, indicating low structural signal-to-noise ratio and reduced threshold stability. 

Such instabilities underscore the difficulty of designing fixed alert thresholds and further motivate multi-layer extensions that can incorporate sentiment or macro shocks when correlation structure alone is insufficient.

\subsection{Interpretable Early-Warning Signals}
Finally, the temporal progression of risk scores offers a complementary perspective to aggregate metrics: while AUROC summarizes discrimination at a single threshold, score trajectories illuminate how model confidence accumulates or dissipates over time, providing an interpretable early-warning signal for practitioners. This temporal view enables risk managers to track not just whether a warning is issued, but how the underlying risk assessment evolves leading up to stress events.

\section{Interpretability and Attribution Analysis}
\label{sec:interpretability}

\subsection{Feature Attribution Across Crisis Regimes}
For GNN models, we use attention-based mechanisms or gradient sensitivity analysis for interpretability. The snapshot GNN reveals which nodes and connections dominate during stress periods. Nodes corresponding to highly liquid sector ETFs receive elevated attention weights, suggesting the model focuses on broad market structure rather than idiosyncratic volatility.

From a systemic-risk perspective, interpretability serves two purposes. First, practitioners can identify which market components contribute most strongly to model warnings, providing transparency for risk-governance workflows. Second, it helps distinguish structurally meaningful signals from spurious short-term noise. As SRR incorporates sentiment and multi-layer dependencies, attention patterns may reveal cross-layer contagion pathways, improving diagnostic power of early-warning tools.

\subsection{Cross-Layer Attribution Patterns}
Interpretability patterns differ across crisis regimes. During Dot-com and GFC, feature and attention attributions emphasize broad structural factors: average correlation, volatility clustering, and sector co-movement. These align with gradual systemic buildup. In contrast, attribution patterns in the COVID–19 period appear diffuse and inconsistent, reflecting the absence of structural precursors. 

This contrast highlights the value of interpretability not only for understanding model behavior when the models succeed but also for diagnosing the structural blind spots of correlation-only GNNs. As additional graph layers are incorporated into SRR, attribution tools can reveal cross-layer contagion pathways and help identify whether fragility is driven by structural, behavioral, or macroeconomic components.

\subsection{Future Directions for Explainability}
We leave a full quantitative study of temporal attention patterns and multi-layer graph attributions to future work, once the complete SRR model with sequence modelling has been trained across all crisis windows. Key areas for investigation include:

\begin{itemize}
    \item Temporal attention weights to identify critical transition points
    \item Cross-layer importance analysis to determine which graph layers dominate in different crisis types
    \item Node-level attribution to identify systemically important institutions
    \item Edge-level analysis to track contagion pathways
\end{itemize}

\end{document}